\documentclass[12pt]{article}

\usepackage{amsmath, latexsym}
\usepackage[psamsfonts]{amsfonts}
\usepackage{amssymb}
\usepackage[authoryear]{natbib}

\newif\ifpdf
\ifx\pdfoutput\undefined
   \pdffalse
\else
   \pdfoutput=1
   \pdftrue
\fi

\ifpdf
    \usepackage{color}
    \definecolor{myred}{rgb}{0.5,0,0}
    \definecolor{myblue}{rgb}{0,0,0.75}
    \definecolor{mygreen}{rgb}{0,0.5,0}
   \usepackage[pdftex,
               pdftitle={Credit Risk Contributions to Value-at-Risk},
               pdfsubject={CreditRisk+ model},
               pdfkeywords={CreditRisk+; Value-at-Risk (VaR); risk contribution; conditional expectation},
               pdfauthor={Alexandre Kurth; Dirk Tasche},
               pdfstartview=FitBH,
               breaklinks=true,
               colorlinks=true,
               citecolor=myred,
               linkcolor=myblue,
               urlcolor=mygreen]{hyperref}
\else
    \newcommand{\href}[1]{}
\fi

\def\VaR{{\mathit{VaR}}}     
\def\ES{{\mathit{ES}}}       
\def\EL{{\mathit{EL}}}       
\def\UL{{\mathit{UL}}}       

\renewcommand{\section}{\subsection}

\begin{document}
\thispagestyle{empty} \begin{flushright} {\scriptsize November 22,
2002}
\end{flushright}

\begin{center}
{\LARGE\bf Credit Risk Contributions to Value-at-Risk}

\vspace{3mm}
{\LARGE\bf and Expected Shortfall}
\end{center}

\vspace{10mm}

\begin{tabular}{ll}

\vspace{3mm}

{\sc Alexandre Kurth}, & {\it UBS AG} \footnotemark\ \\

\vspace{3mm}

{\sc Dirk Tasche}, & {\it Deutsche Bundesbank} \footnotemark\ \\
\addtocounter{footnote}{-1}
\end{tabular}

\footnotetext{Credit Risk Control, UBS Wealth Management \&
Business Banking, P.O.\ Box, 8098 Zurich, Switzerland. email:
alexandre.kurth@ubs.com}\addtocounter{footnote}{1}
\footnotetext{Deutsche Bundesbank, Postfach 10 06 02, 60006
Frankfurt a.\ M., Germany. email: tasche@ma.tum.de}

\vspace{5mm}

\begin{abstract}
\noindent This paper presents analytical solutions to the problem
of how to calculate sensible $\VaR$ (Value-at-Risk) and $\ES$
(Expected Shortfall) contributions in the CreditRisk$^+$
methodology. Via the $\ES$ contributions, $\ES$ itself can be
exactly computed in finitely many steps. The methods are
illustrated by numerical examples.
\end{abstract}

\section{Introduction}\label{intro}

From a theoretical point of view, there is no need to assign risk
contributions to the parts of a portfolio. The reason is that no rational
investor would be happy with a sub-optimal portfolio in the sense of best
possible return-risk ratios. Therefore the assumption of already optimized
portfolios predominates in the scientific literature. Of course, in case of an
optimal portfolio surveillance of sub-portfolios is dispensable since by
definition such a portfolio cannot be improved.

Nevertheless, in practice, and in particular in credit practice, optimal
portfolios are quite rare. Even if the causes of bad performance in a credit
portfolio have been located, as a consequence of limited liquidity it will in
general be impossible to optimize the portfolio in one step. Indeed, credit
portfolio management is rather a sequence of small steps than a one strike
business. However, steering a portfolio step by step requires detailed risk
diagnoses. The key to such diagnoses is assigning appropriate risk
contributions to the sub-portfolios.

In a recent paper, \citet{KS02} have reviewed several approaches
to the problem of how to define sensible risk contributions. We
investigate here the ``continuous marginal contribution'' approach
for the CreditRisk$^+$ \citep{CR97} credit portfolio model. More
exactly, we determine analytically the contributions of each
obligor to $\VaR$ (Value-at-Risk) and $\ES$ (Expected Shortfall).
For continuous distributions $\ES$ is the conditional tail
expectation. Moreover, $\ES$ is coherent \citep[cf.][]{ADEH99},
i.e., in contrast to $\VaR$, expected shortfall always respects
diversification as expressed by the subadditivity axiom.

This paper shows that the composition of tail losses may
significantly differ when measured by $\ES$, or $\VaR$, or (a
multiple of) the standard deviation. This fact has implications on
the performance measurement for an active portfolio management.
Apart from this, the contribution to $\ES$ is a useful indicator
for concentrations, as shown in the examples (Section
\ref{several_segments}). A decreasing list of $\ES$-contributors,
even for large portfolios, may reveal that not just the largest
exposures contribute the most, but also certain low quality loans
in particularly volatile segments. This is a valuable input in
analyzing the composition of tail losses, especially in terms of
single name stress losses.

It turns  out that the calculation of the $\VaR$- and $\ES$-contributions
requires per segment only one extra run in the CreditRisk$^+$ model. As a
byproduct, we show that in the CreditRisk$^+$ model $\ES$ can be calculated by
only making use of the loss distribution up to the corresponding quantile
loss.

\citet{Y01a, Y01b} observed that due to estimation errors
simulation-based computations of $\VaR$, $\ES$, and of their
respective risk decompositions must be rather extensive in order
to arrive at acceptable precision. The analytical results
presented in this paper show that CreditRisk$^+$ offers a fast
alternative not only with respect to $\VaR$ and $\ES$ themselves
but also for their risk decompositions.

This article first outlines the main features of the CreditRisk$^+$ model. In
Section \ref{measure} we introduce some aspects of risk measures. The formulas
for the $\VaR$- and $\ES$-contributions are derived in Sections
\ref{VaR_Contributions} and \ref{ES_Contributions} for the one segment case,
and are then extended to several segments in Section \ref{several_segments}.
The paper finishes with an example where the different risk measures are
compared based on independent, uncorrelated segments, as well as correlated
segments. The appendix sketches the proof of the main statement given by
Equation (\ref{eq:58}).

\section{Short Overview of CreditRisk$^+$}\label{overviewCR+}

We will briefly outline the main aspects and relations of the
CreditRisk$^+$ model \citep{CR97} in the one segment case. The
starting point of this credit portfolio model is the following
equation for the random portfolio loss $L$ over all obligors $A$:
\begin{equation}\label{loss_variable}
L\,=\, \sum_A I_A \,\nu_A,
\end{equation}
where $I_A$ is interpreted as an indicator variable describing the default
event of $A$, i.e.\ $I_A=1$ if $A$ defaults and $I_A=0$ otherwise, and
$\nu_A$ is the exposure net of recovery%
\footnote{In this paper, the loss severities are assumed to be constant for
each obligor. For an extension to variable severities confer
\citet{BKW01}.}%
.\ %
Let us denote the (unconditional) default probability of obligor $A$ by $p_A$,
i.e.\ $p_A$ is the mean of $I_A$. The dependence between obligors is
incorporated by a common risk factor $S$ which will later be assumed to be
gamma-distributed. The mean of the variable $S$ is equal to $\mu = \sum_A
p_A$, and its volatility is denoted by $\sigma$. Then
conditional on $S$, the expectation%
\footnote{%
$S\,p_A / \mu$ may also be regarded as the conditional default probability of obligor
$A$ given state $S$ of the economy. However, in the model under consideration,
$S\,p_A / \mu$ may take values greater than one. Thus, the interpretation
as probability has to be understood in an approximative sense.}
of $I_A$ is $S\,p_A /\mu$.
This default scaling factor $S$ reflects the intensity of the number of
default events in the economy; cf.\ \citet{BKWW99}. The expected loss $\EL$
and the standard deviation $\UL$ of the loss variable $L$ can now be derived
\citep[cf.][]{BKW01}:
\begin{equation}\label{EL-UL-formula}
\EL =\sum_A p_A\, \nu_A, \quad \UL^2 =\bigl(
\frac{\sigma}{\mu}\bigr) ^2 \EL^2 + \sum_A \Big[ p_A
-\bigl(1+\bigl( \frac{\sigma}{\mu}\bigr) ^2 \bigr) p_A^2\Big ]
\,\nu_A^2
\end{equation}
There is a slight difference to the formula in \citet[][Eq.\
118]{CR97}. This is caused by the assumption made there that the
default is modelled by a conditionally Poisson distributed random
variable while here defaults are Bernoulli events\footnote{The
derivation of the $\UL$-formula in \citet{CR97} makes use of
probability generating functions as the exposures are banded.
However, this is not necessary as shown in \citet[][Theorem
1]{BKW01}.}. However, in order to get a hand on the loss
distribution, and not just the first and second moments, it is
convenient to adopt the Poisson approximation and  to make use of
the probability generating function\footnote{For an introduction
of probability generating functions and their properties see
\citet{Fi63}.} $G(z)$ which is defined as a power series of the
form
\begin{equation}
  G(z) = \sum_{n=0}^\infty p(n)\,z^n.
\end{equation}
Here $p(n)$ is the probability of losing the amount $n$ and $z$
is a formal variable. In order to compute the coefficients $p(n)$
we make the following assumptions: First, the exposures (net of recovery) are banded%
\footnote{I.e., the exposures (net of recovery) $\nu_A$ can be assumed to be
integer values.}. Second, the default events $I_A$, conditional on $S$, are
approximated by independent Poisson variables, and third,  the common risk
factor $S$ is gamma-distributed with mean $\mu=\sum_A p_A$ and volatility
$\sigma$.

With these assumptions, it turns out that the probability generating function
of the portfolio loss satisfies the following relation (\citealp[cf.][Eq.\
68]{CR97}, \citealp[and][Eq.\ 13]{BKW01}).
\begin{equation}
\label{gen_func} G(z) = \left({1 - \frac{\sigma^2}{\mu}
\,(Q(z)-1)}\right)^{-\alpha},
\end{equation}
where $\alpha = \mu^2/\sigma^2$, and $Q(z)=\mu^{-1}\sum_A p_A z^{\nu_A}$
denotes  the so-called \emph{portfolio polynomial}. Equation (\ref{gen_func})
can be used to derive a recursive condition on the coefficients $p(n)$ which
is due to \citet{Panjer80, Panjer81} (\citealp[cf.\ also][Eq.\ 77]{CR97},
\citealp[and][Eq.\ 14]{BKW01}):
\begin{equation}\label{rec_formula}
p(n) = \frac{1}{n\,(1+\sigma^2/\mu)}
           \sum_{j=1}^{\min(m, n)} \bigl(\sum_{A:\,\nu_A=j}p_A\bigr)
           \bigl\{{\alpha}^{-1} n +
\bigl(1-\alpha^{-1}\bigr)\,j\bigr\}\, p(n-j),
\end{equation}
where $m=\deg(Q)$ is the largest exposure in the portfolio, and
the initial value is given by
$p(0)=(1+\sigma^2/\mu)^{-\alpha}$.

These equations can be extended to the case of several segments,
i.e.\ to the case where the economy is described by a set of systematic risk
factors $S_1,\ldots,S_N$. More specifically, under the
assumption of independent segments with gamma-distributed factors, the extension for
(\ref{rec_formula}) is given in \citet[][Eq.\ 77]{CR97}.

\section{How to Measure Risk}\label{measure}
A risk measure is a metric measuring the uncertainty of the
portfolio loss. If a portfolio is given by an element of the set
$P:=\{ \nu =(\nu_A \mid A \text{ obligor}) \}$ where $\nu_A$
represents the exposure net of recovery of $A$, then the portfolio
loss is given by (\ref{loss_variable}), i.e. $L = L(\nu )= \sum_A
I_A\,\nu_A$. Formally, a risk measure is described by a function
$\rho : P \to \mathbb{R}$, and for every portfolio $\nu \in P$,
the number $\rho(\nu)$ is the risk of $\nu$.

There are two particularly popular examples of risk measures. One
is the \emph{standard deviation} (or \emph{unexpected loss},
$\UL$),
\begin{equation}\label{eq:50}
  \UL(\nu) \ =\ \sqrt{\mathrm{E}\bigl[\bigl(L(\nu) -
  \mathrm{E}[L(\nu)]\bigr)^2\bigr]},
\end{equation}
where $L(\nu)$ is defined by (\ref{loss_variable}). The other is
the \emph{value-at-risk} ($\VaR$) at level $\delta$, defined as
the $\delta$-quantile of the portfolio loss $L(\nu)$:
\begin{equation}\label{eq:51}
  \VaR_\delta(\nu) \ = \ q_\delta(L(\nu))\ = \ \min\bigl\{ l \in
  \mathbb{R} : \Pr\bigl[L(\nu) \le l\bigr] \ge \delta \bigr\}.
\end{equation}
$\UL$ is popular mainly for its computational simplicity. The
level $\delta$ of $\VaR$ has an immediate interpretation as
probability of solvency of the lender. However, both of these risk
measures exhibit certain counter-intuitive properties
\citep[cf.][]{ADEH99}. $\UL$ is not monotonous, i.e.\ it may
happen that $L(\nu') \le L(\nu)$ with probability 1 but $\UL(\nu')
> \UL(\nu)$. Moreover, $\UL$ does not distinguish the upside and
downside potentials of portfolios.

The problem with $\VaR$ is that it is in general not sub-additive.
There may occur portfolios $\nu$ and $\nu'$ such that $\VaR(\nu +
\nu') > \VaR(\nu) + \VaR(\nu')$. Translated into economic terms,
this means that a $\VaR$-investor may face situations where
diversification does not pay.

The most promising alternative to $\VaR$ seems to be
\emph{Expected Shortfall} ($\ES$) (or \emph{Conditional
Value-at-Risk (CVaR)}, \citealp[see][]{AT01}, and \citealp{RU01}).
From its definition
\begin{equation}\label{eq:52}
  \ES_\delta(\nu) \ = \ (1-\delta)^{-1} \int_\delta^1
  q_u(L(\nu))\,d u
\end{equation}
it is clear that $\ES$ dominates $\VaR$ and hence preserves the
interpretation in terms of solvency probability. More important,
$\ES$ is fully coherent in the sense of \citet{ADEH99}. In
particular, $\ES$ is sub-additive.

It can be shown \citep{AT01} that $\ES_\delta$ is a convex
combination of $\VaR_\delta(\nu)$ and the conditional expectation
of $L(\nu)$ given that $L(\nu) > q_\delta(L(\nu))$. As a
consequence, $\ES_\delta$ will be very well approximated by
$\mathrm{E}[L(\nu) \,|\,L(\nu) > q_\delta(L(\nu))]$ if the
probability $\Pr[L(\nu) = q_\delta(L(\nu))]$ is small. This will
be the case in particular in situations like those considered in
the paper at hand. Here $L(\nu)$ represents the loss in a big loan
portfolio. Therefore, in the sequel, we will make use of the
conceptually simpler but slightly unprecise respresentation
\begin{equation}\label{ES_def}
  \ES_\delta(\nu)\,=\, \mathrm{E}[L(\nu) \,|\,L(\nu) >
  q_\delta(L(\nu))].
\end{equation}

\section{Contributions to Value at Risk}\label{VaR_Contributions}
For risk measurement and management it is crucial to allocate the risk to
individual obligors or groups of obligors. So, after the decision for a risk
measure, a further decision concerning the risk allocation is necessary.
CreditRisk$^+$ defines the contribution $C_A^{(\UL)}$ of obligor $A$ to the
portfolio $\UL$ by
\begin{equation}\label{eq:55}
  C_A^{(\UL)} = \nu_A \frac{\partial \UL(\nu)}{\partial \nu_A} =
  \frac{\nu_A}{2\, \UL} \frac{\partial \UL^2(\nu)}{\partial \nu_A} =
  \frac{\nu_A\,\mathrm{cov}[I_A, L]}{\UL}.
\end{equation}
This is a transparent and fast way of allocating risk to
individual obligors. Note that the $\UL$ contributions sum up to
$\UL$, i.e.\ $\sum_A C_A^{(\UL)} = \UL$. Actually, to define the
contributions to portfolio risk by partial derivatives is the only
way which is compatible to portfolio optimization \citep[][Theorem
4.4]{T99}.

Unfortunately, $\VaR$ is not in general differentiable with
respect to the exposures. However, as several authors
\citep{H99,GLS00,L99,T99} noticed, in case when $\VaR$ is
differentiable, its derivative coincides with an expression that
exists always under the assumptions of the CreditRisk$^+$ model,
namely
\begin{equation}\label{eq:56r}
  \frac{\partial \VaR_\delta(\nu)}{\partial \nu_A}\ = \
  \mathrm{E}[I_A\, \mid\, L(\nu) = q_\delta(L(\nu))].
\end{equation}
$\mathrm{E}[I_A\, \mid\, L(\nu) = q_\delta(L(\nu))]$ is the
conditional expectation of the Poisson variable $I_A$ given that
the portfolio loss $L(\nu)$ assumes the worst case value
$q_\delta(L(\nu)) = \VaR_\delta(\nu)$. This observation suggests
the definition of
\begin{equation}\label{eq:57}
  C_A^{(\VaR_\delta)} \ = \ \nu_A \,\mathrm{E}[I_A\, \mid\, L(\nu) = q_\delta(L(\nu))]
\end{equation}
as the contribution of obligor $A$ to portfolio $\VaR$ in the
general case. As with $\UL$, also with Definition (\ref{eq:57}) of
the $\VaR$ contributions the additivity property holds, i.e. \
$\sum_A C_A^{(\VaR_\delta)} = \VaR_\delta$.

Under the assumptions of Section \ref{overviewCR+} that have led
to (\ref{gen_func}), one can show \citep[see Appendix for a proof
in the one segment case and][for a different, general
proof]{Tasche02} that the expectation on the right-hand side of
(\ref{eq:57}) is given by
\begin{equation}\label{eq:58}
\mathrm{E}_\alpha[I_A\, \mid\, L = q_\delta(L)]\ =\ p_A
\frac{\Pr_{\alpha+1}[L = q_\delta(L)-\nu_A]}{\Pr_{\alpha}[L =
q_\delta(L)]},
\end{equation}
where subscript $\alpha$ means that expectation and probability
stem from the original generating function (\ref{gen_func})
whereas subscript $\alpha +1$ means that the probability has to be
derived from (\ref{gen_func}) with $\alpha=\mu^2/\sigma^2$
replaced by $\alpha + 1 = \mu^2 / \sigma^2+1$.

Recall that $\mathrm{E}_\alpha[I_A\, \mid\, L = q_\delta(L)]$ in
(\ref{eq:58}) is an approximation for the probability of ``obligor
A has defaulted'' conditional on the event that the sum of losses
equals $q_\delta(L)$. By the definition of conditional
probability, $\Pr_{\alpha+1}[L = q_\delta(L)-\nu_A]$ therefore can
be interpreted as an approximation of the probability of ``sum of
losses equals $q_\delta(L)$'' conditional on ``obligor A has
defaulted''. Note that $\mathrm{E}_\alpha[I_A\, \mid\, L =
q_\delta(L)]=0$ can happen, see \citet{Tasche02} for a detailed
discussion of this case. In Section \ref{several_segments}, the
formula for the multi-segment case is provided.

Equations (\ref{eq:57}) and (\ref{eq:58}) show that in
CreditRisk$^+$ the $\VaR$ contributions can be calculated by
running the Panjer algorithm (see (\ref{rec_formula})) twice: once
with $\alpha = \mu^2 / \sigma^2$ and again with $\alpha+1 = \mu^2
/ \sigma^2+1$. The set up of this second run is given by scaling
all $p_A$ with the factor $1+\alpha^{-1}$, and the default
volatility $\sigma$ with $\sqrt{1+\alpha^{-1}}$.

\citet{MBT01} took (\ref{eq:56r}) as point of departure for another approach
to $VaR$ contributions by saddlepoint approximation.
In the CreditRisk$^+$ setting, the technique used to arrive at (\ref{eq:58})
yields simple formulas also for this approach \citep{Tasche02}.

\section{Contributions to Expected Shortfall}\label{ES_Contributions}
Via the decomposition
\begin{equation}\label{decomp}
\mathrm{E}[L(\nu) \,|\,L(\nu) >
  q_\delta(L(\nu))] \,=\,\sum_A \nu_A\,\mathrm{E}[I_A \,|\,L(\nu) >
  q_\delta(L(\nu))]
\end{equation}
Equation (\ref{ES_def}) suggests the definition of $\ES$
contributions as
\begin{equation}\label{ES_contrib}
C_A^{(\ES_\delta)} \ =\ \nu_A\,\mathrm{E}[I_A \,|\,L(\nu) >
  q_\delta(L(\nu))].
\end{equation}
This definition can be justified in a similar way as the
definition of $C_A^{(\VaR_\delta)}$ in (\ref{eq:57}).

From (\ref{eq:58}) we know how to calculate
$\mathrm{E}[I_A\,\mid\,L(\nu) = q_\delta(L(\nu))]$ in the
CreditRisk$^+$ model. It is not hard to see that a similar formula
obtains for $\mathrm{E}[I_A\,\mid\,L(\nu) > q_\delta(L(\nu))]$,
namely
\begin{equation}\label{eq:63}
  \mathrm{E}_\alpha[I_A\, \mid\, L > q_\delta(L)]\ =\ p_A
\frac{\Pr_{\alpha+1}[L > q_\delta(L)-\nu_A]}{\Pr_{\alpha}[L >
q_\delta(L)]},
\end{equation}
where the subscripts $\alpha$ and $\alpha + 1$ have the same meanings as in
(\ref{eq:58}). Note that (\ref{eq:63}) can be calculated in finitely many
steps since $\Pr_{\alpha+1}[L > q_\delta(L)-\nu_A] = 1 - \Pr_{\alpha+1}[L \le
q_\delta(L)-\nu_A]$ and $\Pr_{\alpha}[L > q_\delta(L)] = 1 - \Pr_{\alpha}[L
\le q_\delta(L)]$. This observation can be transferred to $\ES$ itself: The
determination of the actual expected shortfall by the recursion-equation
(\ref{rec_formula}) requires that infinitely many elements are calculated. But
applying the sum of all contributions $\sum_A C_A^{(\ES_{\delta})}$ using
Equation (\ref{eq:63}) yields the expected shortfall precisely without knowing
the tail beyond the corresponding $\delta$-quantile loss.

\section{Extension to several segments}\label{several_segments}
There are two possibilities to extend the concept of the contributions to
expected shortfall (resp.\ value-at-risk) to several segments. First, one can
make use of $N$ independent segments described by an $N$-tuple of independent
systematic risk factors $S_1,\ldots,S_N$ which are gamma distributed. In this
case one has to introduce factor loadings $\omega_{Aj}$ to be the portions of
the default probability $p_A$ allocated to segment $j$. This means that
$\omega_{A0}= 1-\sum_{j=1}^N \omega_{Aj}$ is the specific weight of $A$
accounting for the idiosyncratic default risk. And, conditional on
$S_1,\ldots,S_N$, the expectation of $I_A$ is given by
\begin{equation}\label{mult_loading}
  \mathrm{E}[I_A\,|\,S_1,\ldots,S_N] \,=\,
  p_A\,\frac{\omega_{A 0}\,\mu_0 + \sum_{j=1}^N \omega_{A j}\,S_j}{\sum_{j=0}^N \omega_{A
  j}\,\mu_j},
\end{equation}
with $\mu_j > 0$, $j=1,\ldots, N$, being the unconditional expectation of
$S_j$ and $\mu_0 \ge 0$ being constant.

The extension of the $\VaR$-contribution, as in Equation
(\ref{eq:58}), to $N$ independent segments has the following form:
\begin{equation}
  \label{VaR_contrib_mult}
\mathrm{E}_\alpha[I_A\,|\, L = q_\delta(L)] \, = \,p_A\,
\frac{\displaystyle{\sum_{j=1}^N \omega_{Aj}\, \Pr\nolimits_{\alpha(j)}} [ L =
q_\delta(L) - \nu_A] + \omega_{A0}\,\Pr\nolimits_{\alpha}[ L =
q_\delta(L)-\nu_A]} {\Pr_{\alpha}[ L = q_\delta(L)]},
\end{equation}
where $\alpha(j) =(\alpha_1,\ldots,\alpha_j +1, \ldots, \alpha_N)$. The
expression $\Pr_{\alpha(j)}[ L = q_\delta(L) - \nu_A]$ is obtained by
evaluating the corresponding element of the multivariate recursion of Equation
(\ref{rec_formula}) \citep[cf.][Eq.\ 77  for $N$ independent segments]{CR97}
where the $j$-th gamma distribution is specified by the pair
$(\alpha_j+1,\beta)$.

Similarly, by replacing the equality signs in $\mathrm{P}_{\alpha(j)}$ with
inequalities $>$ one obtains the corresponding  formula for the contribution
to $\ES$ for $N$ independent segments.

However, a segmentation with independent segments is rather
restricting. A dependence structure between the obligors is
achieved by apportioning the default probabilities to the
orthogonal factors as done with the $\omega_{Aj}$. This is rather
arbitrary, and therefore creates an additional source of
uncertainty. These factors typically represent the default
behavior of industrial sectors or geographical areas etc., and as
such mostly incorporate significant correlations between them.

To get around the restriction of independent segments one can
introduce the correlation matrix of the $N$ gamma variables
$S_1,\ldots,S_N$, and can match the $\UL$-formula of the loss
distribution for these $N$ segments with the corresponding one
segment $\UL$-formula (see Equation (\ref{EL-UL-formula})). This
procedure is introduced in \citet[][Eq.\ (13)]{BKWW99}. Now, the
portfolio is unified to one segment, and the contributions to
$\VaR$ and $\ES$ are determined as in the previous sections
following Equations (\ref{eq:58}) resp.\ (\ref{eq:63}). This
method is simple and fast. However, it destroys information about
dependence and the contributions tend to the middle since the
matched dependence structure is given by one volatility which
turns out to be some average of the segment volatilities.

Orthogonalization of segment correlations does not represent an alternative
because the process of orthogonalization is not unique. It can be shown by
examples that significant differences in $\VaR$, $\ES$ and their contributions
show up.

\section{Numerical Example}\label{examples}
We want to show that the risk contributions according to the
expected shortfall $\ES$ and value-at-risk $\VaR$ may
significantly differ from the risk contributions according to the
standard deviation. We will do so by using a wholesale bank
portfolio that includes a retail subportfolio as well as
commercial loans of various sizes. The following table describes
the portfolio which is characterized by the loan sizes given in
the first line.

\begin{center}{\scriptsize
\begin{tabular}{|r|rr|rrrrrr|}  \hline
 & \multicolumn{2}{c|}{{\bf segment 1}}  & \multicolumn{6}{c|}{{\bf segment 2}} \\
\hline
Exposure per obligor [in Mio CHF] & 1 & 1 & 10 & 20 & 100 & 500 & 1'000 & 2'000 \\
\hline
\# obligors & 10'000 & 10'000 & 1'000 & 500 & 100 & 10 & 2 & 1 \\
 PD for each obligor & 0.5\% & 1\% & 1\% & 1.75\%
& 1.75\% & 1.25\% & 0.70\% & 0.30\% \\ \hline Exposure in class
[in Mio CHF] & 10'000 & 10'000 & 10'000 & 10'000 & 10'000
& 5'000 & 2'000 & 2'000\\
Exposure in \% of total & 16.9\% & 16.9\% & 16.9\% & 16.9\% &
16.9\% & 8.5\% & 3.4\% & 3.4\%
\\ \hline
\end{tabular}
} \nopagebreak

\nopagebreak \refstepcounter{table} {\scriptsize
\begin{tabular}{l}
Table \thetable\label{table:portfolio_description}: Specification of the
sample portfolio. PD means ``probability of default''.
\end{tabular}}
\end{center}

We assume that all loan exposures are given net of recovery. As
indicated in the table we further assume a segmentation with two
segments: The loans of exposure $1$ Mio CHF form segment $1$ (i.e.
the retail segment), and all others build the second segment
(commercial segment).

Using the portfolio in Table \ref{table:portfolio_description}, we
are going to present three different cases for the computation of
$\ES$, $\VaR$, $\UL$ and the corresponding contributions. The
first case is characterized by the segment volatilities $\sigma_1,
\sigma_2$ of default rates and independence of segments $1$ and
$2$, where we use Equation (\ref{VaR_contrib_mult}) for
determining the $\VaR$- (resp.\ $\ES$-) contributions. In the
second run we take the same segment volatilities as in the first
case, but now we use the method of moment matching\footnote{Here
we assume zero correlation whereas in case 1 we assume
independence. Note that uncorrelatedness of two gamma
distributions does not imply their stochastic independence.} as
described in the last paragraph of Section \ref{several_segments}.
In the last case we even introduce correlations between both
segments and use again the moment matching method to calculate the
contributions (see Table \ref{table:portfolio_numbers}). Note that
the covariance of 0.21 in the last case corresponds to a $70\%$
correlation between the segments. The segment volatilities
$\tilde{\sigma}_i$ are given in terms of their relative means;
e.g., the volatility $\sigma_1$ of defaults in segment $1$ for all
three cases is $\tilde{\sigma}_1=\sqrt{0.16}$ times the expected
number of defaults in segment $1$, hence $\sigma_1 = \sqrt{0.16}
\times 0.75\% \times 10'000$. The following table gives the exact
description for these three cases.

\begin{center}{\scriptsize
\begin{tabular}{|cc|ccccccc|}  \hline
Case name & dependence structure & Total exposure & $\EL$ & $\EL$
in
\%& $\UL$ & $\UL / \EL$ & $\VaR$ & $\ES$\\
\hline independence & {\scriptsize \begin{tabular}{c}
$\tilde{\sigma}_1 = \sqrt{0.16}$\\
$\tilde{\sigma}_2 = \sqrt{0.56}$
\end{tabular}} &
59'000 & 682 & 1.16\% & 490 & 72\% & 2'434 & 2'915\\
\hline zero-correlation & \mbox{{\scriptsize $\left(
  \begin{array}{cc}
    0.16 & 0 \\
    0 & 0.56
  \end{array}\right)$}} &
59'000 & 682 & 1.16\% & 490 & 72\% & 2'357 & 2'805 \\
\hline correlation & \mbox{{\scriptsize $\left(
  \begin{array}{cc}
    0.16 & 0.21 \\
    0.21 & 0.56
  \end{array}\right)$}} &
59'000 & 682 & 1.16\% & 523 & 77\% & 2'481 & 2'954 \\
\hline
\end{tabular}
} \nopagebreak

\nopagebreak \refstepcounter{table} {\scriptsize
\begin{tabular}{l}
Table \thetable\label{table:portfolio_numbers}: Portfolio figures
for three cases of dependence. All absolute numbers are given in
Mio CHF.
\end{tabular}}
\end{center}

Observe that independence does not necessarily lead to the lowest
possible risk measure. The following three tables list the segment
contributions to the measures of risk for the three cases
described above: standard deviation $\UL$, $99$-percentile $\VaR$,
and $\ES$ at $99\%$. In order to compare the contributions we give
the ratios of the relative contributions per exposure class for
the three risk measures.

\begin{center}{\scriptsize
\begin{tabular}{|r|rr|rrrrrr|r|}  \hline
independence & \multicolumn{2}{c|}{{\bf segment 1}}  & \multicolumn{6}{c|}{{\bf segment 2}} & {\bf portfolio totals}\\
\hline
Exposure per obligor & 1 & 1 & 10 & 20 & 100 & 500 & 1'000 & 2'000 & 59'000\\
PD for each obligor & 0.5\% & 1\% & 1\% & 1.75\% & 1.75\% & 1.25\%
& 0.70\% & 0.30\% & \\ \hline
$C^{(\UL)}$ & 3 & 5 & 63 & 113 & 141 & 101 & 37 & 28 & 490 \\
$C^{(\VaR)}$ & 52 & 105 &  282 & 503 & 581 & 434 & 229 &
247 & 2'434\\
$C^{(\ES)}$ & 53 & 105 & 312 & 555 & 643 & 478 & 264 &
504 & 2'915\\
\hline $C^{(\VaR)}\%/ C^{(\UL)}\%$ & 4.15  &  4.15  &  0.91  &
0.89 & 0.83 & 0.87   & 1.26  &  1.78 & \\
$C^{(\ES)}\% /C^{(\VaR)}\%$ & 0.84  &  0.84  &  0.92  &  0.92 &
0.92
&  0.92   & 0.96  &  1.71 & \\
$C^{(\ES)}\% / C^{(\UL)}\%$ & 3.47  & 3.48   & 0.83   & 0.82  &
0.77  &  0.80  &  1.21  &  3.03 &
\\ \hline
\end{tabular}
} \nopagebreak

\nopagebreak \refstepcounter{table} {\scriptsize
\begin{tabular}{l}
Table \thetable\label{table:contrib_ortho1}: Contributions to
various risk measures in case 'independence'. All absolute numbers
are given in Mio CHF.
\end{tabular}}
\end{center}

\begin{center}{\scriptsize
\begin{tabular}{|r|rr|rrrrrr|r|}  \hline
zero-correlation & \multicolumn{2}{c|}{{\bf segment 1}}  & \multicolumn{6}{c|}{{\bf segment 2}} & {\bf portfolio totals}\\
\hline
Exposure per obligor & 1 & 1 & 10 & 20 & 100 & 500 & 1'000 & 2'000 & 59'000\\
PD for each obligor & 0.5\% & 1\% & 1\% & 1.75\% & 1.75\% & 1.25\%
& 0.70\% & 0.30\% & \\ \hline
$C^{(\UL)}$ & 3 & 5 & 63 & 113 & 141 & 101 & 37 & 28 & 490 \\
$C^{(\VaR)}$ & 116 & 233 &  237 & 423 & 499 & 410 & 234 &
205 & 2'357\\
$C^{(\ES)}$ & 123 & 245 & 250 & 447 & 526 & 428 & 262 &
524 & 2'805\\
\hline $C^{(\VaR)}\%/ C^{(\UL)}\%$ & 9.49  &  9.49  &  0.78   &
0.78  & 0.74 &   0.85  &  1.33 & 1.53 & \\
$C^{(\ES)}\% /C^{(\VaR)}\%$ & 0.89  &  0.89  &  0.89  &  0.89  &
0.89 &   0.88  &  0.94 & 2.15 & \\
$C^{(\ES)}\% / C^{(\UL)}\%$ & 8.41  &  8.42  &  0.70  &  0.69  &
0.65  &  0.74  &  1.25 & 3.27&
\\ \hline
\end{tabular}
} \nopagebreak

\nopagebreak \refstepcounter{table} {\scriptsize
\begin{tabular}{l}
Table \thetable\label{table:contrib_ortho2}: Contributions to
various risk measures in case 'zero-correlation'. All absolute
numbers are given in Mio CHF.
\end{tabular}}
\end{center}

\begin{center}{\scriptsize
\begin{tabular}{|r|rr|rrrrrr|r|}  \hline
correlation & \multicolumn{2}{c|}{{\bf segment 1}}  & \multicolumn{6}{c|}{{\bf segment 2}} & {\bf portfolio totals}\\
\hline
Exposure per obligor & 1 & 1 & 10 & 20 & 100 & 500 & 1'000 & 2'000 & 59'000\\
PD for each obligor & 0.5\% & 1\% & 1\% & 1.75\% & 1.75\% & 1.25\%
& 0.70\% & 0.30\% & \\ \hline
$C^{(\UL)}$ & 13 & 26 & 65 & 117 & 143 & 98 & 35 & 27 & 523 \\
$C^{(\VaR)}$ & 128 & 255 &  259 & 462 & 536 & 400 & 211 & 230 & 2'481\\
$C^{(\ES)}$ & 139 & 279 & 284 & 506 & 588 & 444 & 247 & 466 & 2'954\\
\hline $C^{(\VaR)}\%/ C^{(\UL)}\%$ & 2.06  &  2.06  &  0.84  &
0.84  & 0.79  &  0.86  &  1.26  & 1.83& \\
$C^{(\ES)}\% /C^{(\VaR)}\%$ & 0.92  &  0.92  &  0.92  &  0.92  &
0.92  &  0.93  &  0.99  & 1.70 & \\
$C^{(\ES)}\% / C^{(\UL)}\%$ & 1.89  &  1.89  &  0.78   & 0.77  &
0.73 &   0.80  &  1.24  & 3.10&
\\ \hline
\end{tabular}
} \nopagebreak

\nopagebreak \refstepcounter{table} {\scriptsize
\begin{tabular}{l}
Table \thetable\label{table:contrib_ortho3}: Contributions to
various risk measures in case 'correlation'. All absolute numbers
are given in Mio CHF.
\end{tabular}}
\end{center}

Note that in all cases the retail loans (CHF $1$ Mio), and
particularly the large lumpy loans contribute more to Expected
Shortfall $\ES$ (and to $\VaR$) than they do to $\UL$, which shows
that $\ES$ measures concentration more sensitively. For the large
loans this is quite intuitive, since once a tail loss above $\VaR$
is reached, the (lumpy) large loans more often contribute by their
defaults. But apparently, a significantly large amount of small
loans also suffices to produce a large loss. In case 1 the
contributions for $\VaR$ and $\ES$ almost coincide for loans in
segment 1, whereas they increasingly diverge for increasing loan
amounts.

Note that in the 'zero-correlation' case (Table
\ref{table:contrib_ortho2}) the $\UL$-contributions are calculated
based on the corresponding covariance matrix \citep[cf.][Equation
(15)]{BKWW99} whereas the $\ES$- and $\VaR$-contributions are
calculated based on the matched one segment approach. This causes
the large differences in the relative contributions of segment
$1$. Calculating the $\UL$-contributions based on the matched one
segment approach would lead to a ratio in the range of one for
$C^{(\ES)}\% / C^{(\UL)}\%$ in segment $1$. Applying one
volatility for the entire portfolio causes a shift of the
contributions to the 'average'. The exposure classes 1 and 10,
both with $p_A=1\%$ have the same total exposure and very similar
contributions for the cases 'zero-correlation' and 'correlation'.
But in the 'independence' case their contributions differ very
much (e.g.\ 105 vs.\ 312 for $\ES$).

Introducing correlations we see a large increase in the
$\UL$-contributions in segment $1$ whereas they remain quite
stable in the other segment. The $\ES$-contributions increase by
approximately $10\%$, except for both largest exposure classes,
where a decrease of up to more than $10\%$ is realized. The
changes of $\VaR$-contributions are similar but smaller, with the
exception of the $\VaR$-contribution reduction in exposure class
$500$.

\section{Conclusion}\label{conclusion}
We have introduced an analytical approach to calculate
contributions to $\VaR$ and $\ES$ of a credit loss distribution in
the CreditRisk$^+$ framework. The formulas we have derived are
easy to implement. The time consumption of calculating these
contributions for all obligors is the same as for calculating the
loss distribution in the moment matching case of CreditRisk$^+$.

The results show that $\VaR$- and $\ES$-contributions may differ
significantly from $\UL$-contributions, in particular for large
exposures. Moreover, especially $\ES$ is sensitive to
concentrations coming from large loans which represent a source
for stress losses in a loan portfolio. Hence, if a bank wants to actively
manage its loan portfolio whose performance is
measured (among others) by tail losses, it should take into account that such
tail events are composed quite differently when measured by expected shortfall
instead by standard deviation (or a multiple thereof).

\bigskip

\noindent {\bf Acknowledgements.} The authors thank Isa Cakir,
Bernd Engelmann, and Armin Wagner for helpful suggestions and
valuable remarks. Alexandre Kurth works in the unit responsible
for modeling credit risk at ``UBS Wealth Managment \& Business
Banking Division'' of UBS AG. Dirk Tasche has a position in the
banking supervision department of Deutsche Bundesbank. Opinions
expressed in this article are those of the authors and do not
necessarily reflect the opinions of UBS or Deutsche Bundesbank.


\appendix
\section*{Appendix}\label{app}
%
%
We sketch here the proof of Equation (\ref{eq:58}). Denote as usual by
$I_{\{L=t\}}$ the indicator function of the event $\{L=t\}$, i.e.\ $I_{L=t} =
1$ if $L=t$ and  $I_{L=t} = 0$ otherwise. By definition, we have
\begin{equation}\label{eq:a1}
  \mathrm{E}_\alpha[I_A\, \mid\, L = q_\delta(L)]\ =\  \frac{\mathrm{E}_\alpha[I_A\,I_{\{L = q_\delta(L)\}}]}
  {\Pr_\alpha[L = q_\delta(L)]}.
\end{equation}
We will now compute the generating function of the sequence $t \mapsto
\mathrm{E}_\alpha[I_A\,I_{\{L = t\}}]$ which is defined as the function
\begin{equation*}
z \mapsto \mathrm{E}_\alpha[I_A\,z^L]\, =\, \sum_{t=0}^\infty
\mathrm{E}_\alpha[I_A\,I_{\{L = t\}}]\,z^t.
\end{equation*}
Since in the CreditRisk$^+$ model, conditional on the intensity
$S$, $I_A$ is approximated by a Poisson variable with intensity
$\frac{p_A}{\mu}\,S$ and the default events are assumed to be
conditionally independent, we obtain
\begin{align}
 \mathrm{E}_\alpha[I_A\,I_{\{L = t\}}]  & = \sum_{k=0}^\infty k\,
 \Pr\nolimits_\alpha[I_A =k, \sum_{B\not=A} \nu_B\,I_B = t - \nu_ A\, k]\notag\\
   & = \sum_{k=1}^\infty k\,\mathrm{E}_\alpha\bigl[(k!)^{-1}\bigl(\frac{p_A}{\mu}S\bigr)^k
   e^{-\frac{p_A}{\mu}S} \Pr\nolimits_\alpha[\sum_{B\not=A} \nu_B\,I_B = t - \nu_ A\,
   k\,|\,S]\bigr]\notag\\
   & = \frac{p_A}\mu \sum_{k=0}^\infty \mathrm{E}_\alpha\bigl[S
   \,\Pr\nolimits_\alpha[I_A =k\,|\,S]\, \Pr\nolimits_\alpha[\sum_{B\not=A} \nu_B\,I_B = t - \nu_ A\,
   (k+1)\,|\,S]\bigr]\notag\\
   & = \frac{p_A}\mu \,\mathrm{E}_\alpha[S\,I_{\{L = t - \nu_A\}}].\label{eq:calc}
\end{align}
Equation (\ref{eq:calc}) implies
\begin{equation}\label{eq:power}
\mathrm{E}_\alpha[I_A\,z^L]\,=\,\frac{p_A}\mu
\,z^{\nu_A}\,\mathrm{E}_\alpha[S\,z^L].
\end{equation}
Under the assumption of a gamma-distributed factor $S$ and the conditional
Poisson distribution and independence of the variables $I_A$, it turns out
that the expectation $\mathrm{E}_\alpha[S\,z^L]$ can be calculated the same
way as $\mathrm{E}_\alpha[z^L] = G(z)$, the generating function of the loss
$L$. The result of the calculation is quite similar, namely (with the same
notation as in (\ref{gen_func}))
\begin{equation}\label{newgen_func}
\mathrm{E}_\alpha[S\,z^L] \,=\,\mu \left({1 -
\frac{\sigma^2}{\mu}\,(Q(z)-1)}\right)^{-(\alpha+1)}.
\end{equation}
From (\ref{newgen_func}) and (\ref{eq:power}) we conclude
\begin{align}
  \mathrm{E}_\alpha[I_A\,z^L] & = p_A\,z^{\nu_A} \left({1 -
(\sigma^2/\mu)\,(Q(z)-1)}\right)^{-(\alpha+1)}\notag \\
   & = p_A\,\mathrm{E}_{\alpha+1}[z^{L+\nu_A}].
  \label{eq:id}
\end{align}
Recall that both sides of (\ref{eq:id}) can be written as power series. Hence
we can conclude that
\begin{equation}\label{eq:coeff}
  \mathrm{E}_\alpha[I_A\,I_{\{L = t\}}] \,=\, p_A\,
  \Pr\nolimits_{\alpha+1}[L=t-\nu_A]
\end{equation}
for $t = 0, 1, 2, \ldots$. Choosing $t= q_\delta(L)$ now proves (\ref{eq:58}).

\end{document}